\documentclass{icrc29}
\usepackage{graphicx,amssymb,amsmath,times,epsfig,floatflt}

\newcommand{\etal}{{\it et al}}

\begin{document}
%Title of paper
\title[Validation]{Validation of the Real and Simulated Data of
the Pierre Auger Fluorescence Telescopes}
\author[L. Perrone \etal] {
 A.Ewers$^a$, H.Geenen$^a$, K-H Kampert$^a$, L. Perrone$^a$, 
 S. Robbins$^a$, V. Scherini$^a$,
 M Unger$^b$ \newauthor
  for the Pierre Auger Collaboration\\
(a) Universit\"{a}t Wuppertal, Fachbereich Physik, D-42097 Wuppertal,
Germany \\
(b) Forschungszentrum Karlsruhe, Institut f\"{u}r Kernphysik, 76021
    Karlsruhe, Germany\\}
% H.Geenen, K-H Kampert, L. Perrone, 
% S. Robbins, V. Scherini,
% M Unger
% ~\footnote{See the Auger author list published in the same proceedings}
% \newauthor
%  for the Auger Collaboration\\
%}
\presenter{Presenter: L. Perrone (perrone@physik.uni-wuppertal.de), \  
ger-kampert-K-abs3-he14-poster}

\maketitle

\begin{abstract}

The fluorescence detector (FD) of the Pierre Auger Observatory is currently
 operating 18 fluorescence telescopes of the 24 that will be employed in
 the completed detector. These telescopes, grouped in 4 eyes each
 consisting of 6 telescopes, measure the longitudinal profile of cosmic
 ray showers with a 14\% duty cycle.
The reconstruction capability and triggering efficiency have been
 studied using a complete simulation and reconstruction
 production chain, employing both simulated CORSIKA showers and parameterised
 Gaisser-Hillas profiles. The propagation through the atmosphere and
 the detector response are taken into account and simulated in detail.
 These simulated data have been generated in a preliminary analysis
 using the method of importance sampling to efficiently cover the
 energy region of 0.3 - 300 EeV, various shower geometries
 and impact points and different primary particles.
 The distributions of observables have then been
 investigated in both real and simulated data, facilitating the
 validation of the reconstruction and simulation software.
 Comparisons of real and simulated data are discussed and used
 to assess their impact on the data analysis.

\end{abstract}

\section{The CORSIKA simulation sample}

  This paper discusses 
 the performance of the Auger fluorescence telescope, which has been 
 studied with  
a large sample of fully simulated  CORSIKA showers \cite{corsika}.
A detailed description of the Fluorescence Detector 
simulation program is given 
in~\cite{fdsim};
the reconstruction was performed using the Auger Offline software ~\cite{offline}.
To obtain a
sufficiently large number of events the CORSIKA showers have been taken from the
shower database generated in the Lyon computing centre 
for simulation studies with the Auger
detector.
The shower sample consists of 3850 proton showers and 4150 iron showers  
with zenith 
angles of 0$^{\circ}$, 18$^{\circ}$, 26$^{\circ}$, 37$^{\circ}$, 
45$^{\circ}$, 60$^{\circ}$ and energies ranging between 10$^{17.5}$ and 
10$^{20.5}$ eV in steps of 0.25 or 0.5 in the logarithmic scale.   
The CORSIKA showers have been simulated in a slice of 2$^{\circ}$  
in the field of view of  
Bay4 (Los Leones Eye), with uniformly distributed
 core distances. This choice has been made  
in order to optimise the reconstruction and trigger 
efficiency study 
as a function of core distance, rather than simulating the true distribution 
of cosmic ray landing points (uniformly distributed on surface).       
In order to minimise the inefficiency due to low energy showers
landing far away from the eye (with a negligible  
 probability of being triggered), the maximum distance of the generated impact points 
 was to chosen to depend  
 on the shower energy and ranges from 5 km up to a maximum of 60 km. 
The sensitivity of reconstructed energy to the 
atmospheric properties has been investigated by assuming 
two extreme atmospheres with  
aerosol horizontal attenuation lengths at sea level of 12.5 km and 24 km and 
scale height of 2 km. 

%%%%%%%%%%%%%%%%%%%%%%%%%%%%%%%%%%%%%%%%%%%%%%%%%%%%%%%%%%%%%%%%%%%%%%%%%%%%
\section{Trigger efficiency and energy resolution with a given shower geometry}
\label{energyres}

The Pierre Auger Observatory employs two independent detection
techniques, allowing the reconstruction of extensive air showers with
two complementary measurements. Indeed, the combination of information from 
the surface array and the fluorescence telescopes enhances the  
reconstruction capability of these
so called "hybrid" events with respect to the individual
detector components.
%\begin{floatingfigure}[l!]{6.5cm}
%   \includegraphics[width=6.0cm]{revised_vivi/RAD_new.eps}
%\protect\caption{Trigger 
%efficiency as a function of energy for increasing core distances
%ranges.}
%\label{fig:trigger_efficiency}
%\end{floatingfigure}
A description of the hybrid performance of the Pierre Auger Observatory
is given in~\cite{miguel}.\\
In this study, 
the energy resolution of the fluorescence detector has been estimated for 
the case of known fixed shower geometry. 
This assumption is justified by the argument that hybrid reconstruction 
benefits from a more accurate shower geometry with respect to      
the monocular fluorescence reconstruction. 
Setting the geometry to the true value then provides  
a realistic estimate of the energy resolution for the hybrid mode.    
Assumptions for the  atmosphere, detector calibration and fluorescence 
yield calculation have 
been made consistently throughout 
the simulation-reconstruction  chain.
\vspace{-10pt}
\begin{figure}[htb]
\begin{minipage}[t]{6.5cm}
\begin{center}
\epsfig{file=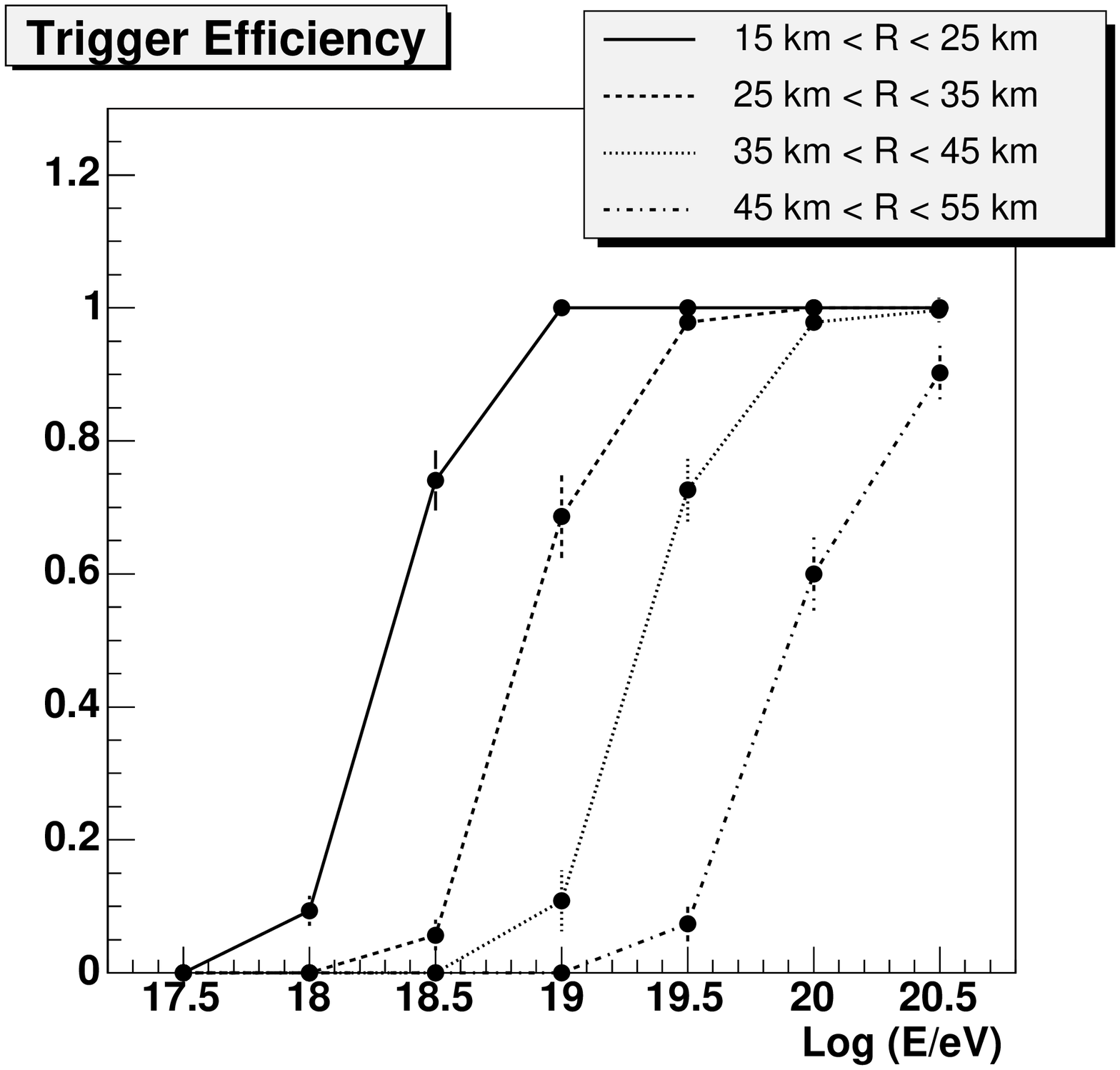, width=6.5cm}
\caption{Trigger efficiency 
as a function of energy for increasing core distances
ranges (all zenith angles merged).}
\label{fig:trigger_efficiency}
\end{center}
\end{minipage}
\hfill
%\vskip -0.3cm
\begin{minipage}[t]{7.9cm}
\begin{center}
\epsfig{file=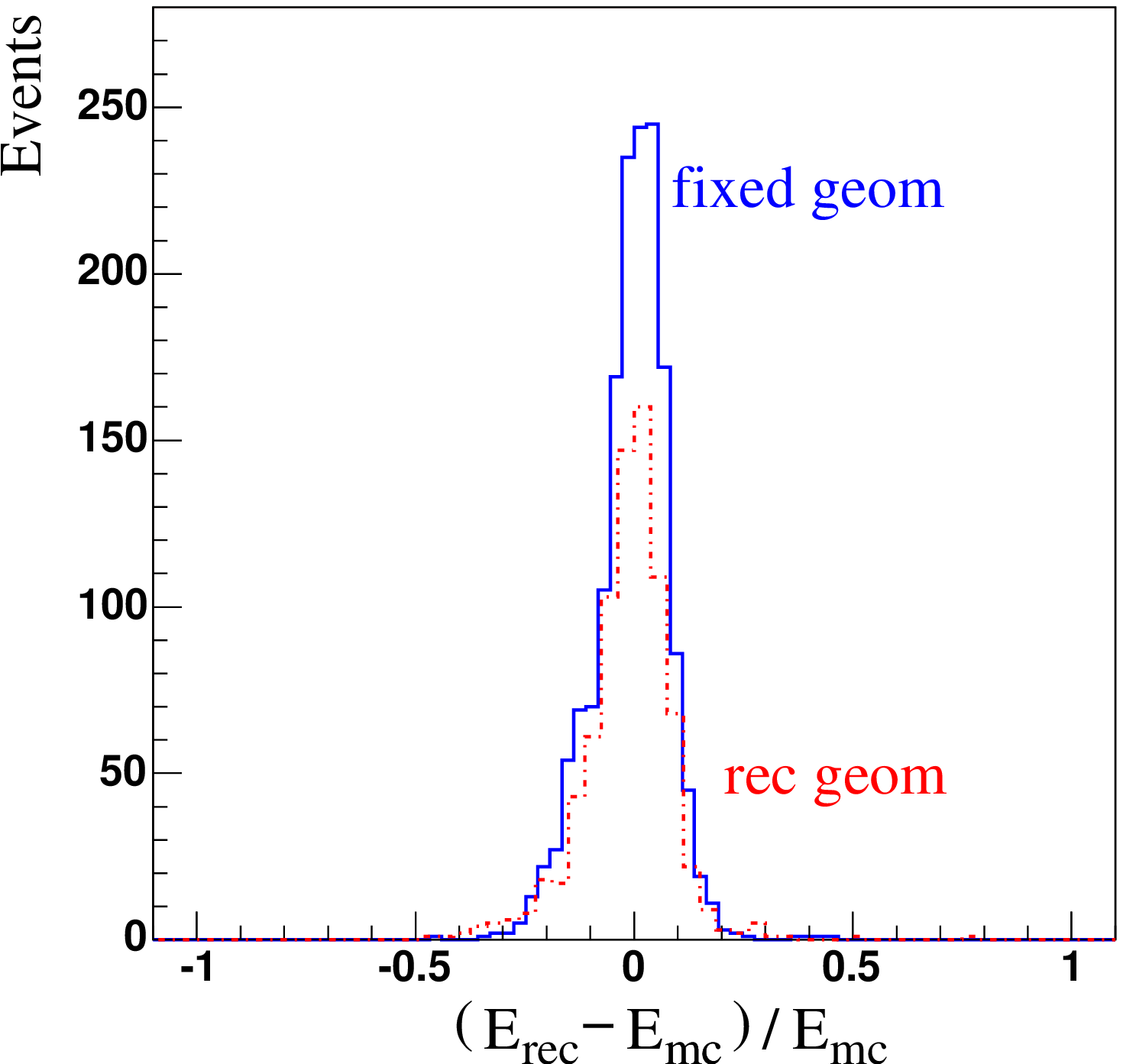, width=7cm}
\caption{
Energy resolution for the simulated data sample
with true geometry (blue line, 1607 events, 
RMS=9\%) and reconstructed monocular geometry (red dot-dashed line, 
798 events, RMS=11\%).}
\label{fig:energy-resolution}
\end{center}
\end{minipage}
\hfill
\vspace{-22pt}
\end{figure}
\begin{figure}[htb]
\begin{center}
\epsfig{file=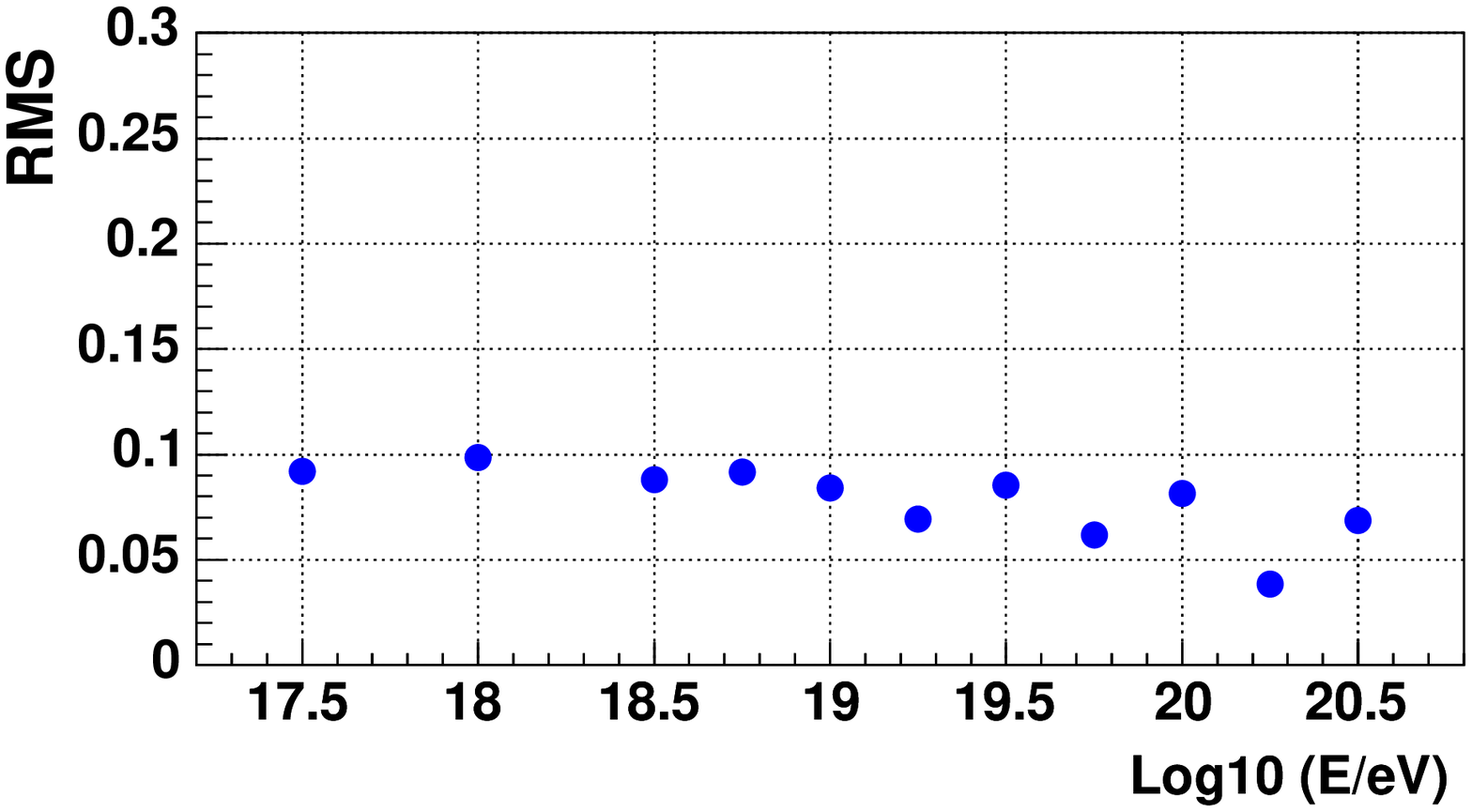,width=5.1cm}
\epsfig{file=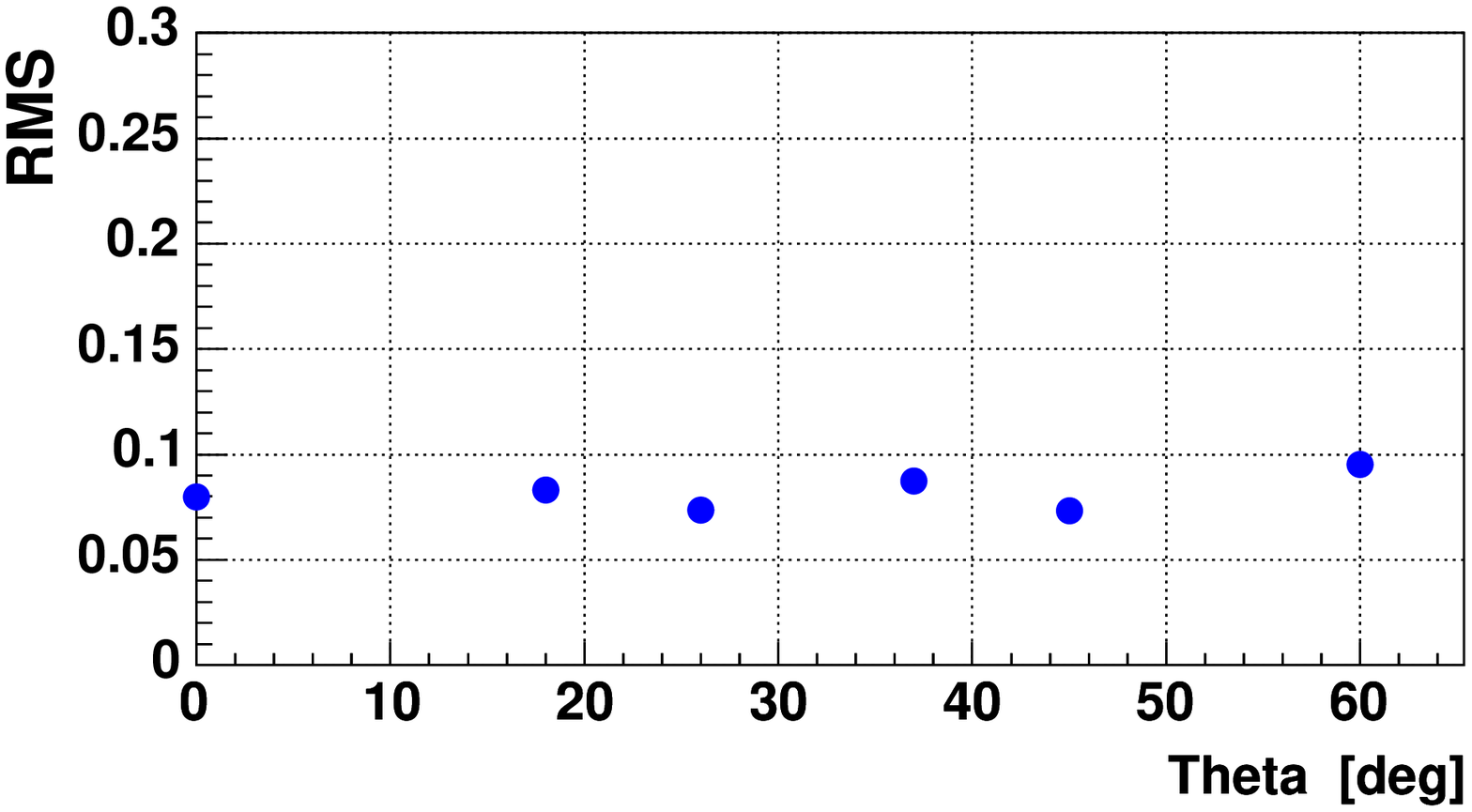,width=5.1cm}
\epsfig{file=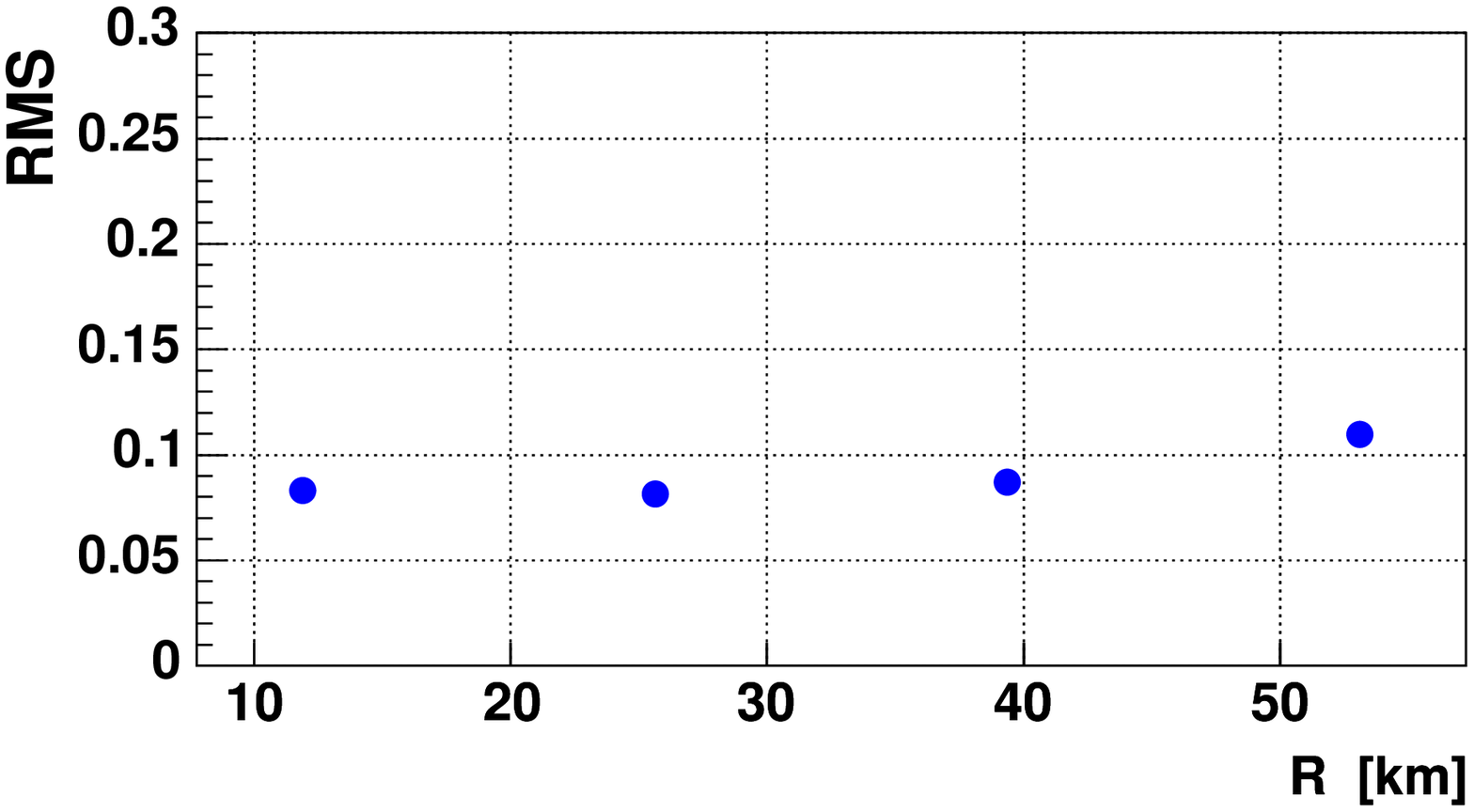,width=5.1cm}
\caption{
RMS of the residual distribution as a function 
of true energy (left), zenith angle (middle) and core distance (right)  
for 
fixed geometry.  
%Statistical errors are smaller than the points.
}
\label{fig:energy-RMS}
\end{center}
%\end{minipage}
%\hfill
\end{figure} 
Fig.~\ref{fig:trigger_efficiency} 
shows the trigger efficiency as a function of energy for 
increasing core distance ranges (all zenith angles merged). 
The trigger efficiency is 100\% up to
a distance of 25 km for showers with energy of 10$^{19}$ eV. A detailed 
calculation of the fluorescence detector aperture for different detector configurations
and using analytical shower profiles (Gaisser-Hillas functions) 
is given in~\cite{petrera}.
The method adopted here for the reconstruction of shower longitudinal profiles
and energies with the Auger Fluorescence telescope is described in~\cite{bellido}.  
In order to focus on "reconstructible" events only,
the observed profile and reconstructed shower depth at maximum ($X_{max}$) 
are required to satisfy the following conditions:\\
- a successful Gaisser-Hillas fit with $\chi^{2}$/Ndof $<$ 5 for the 
 reconstructed longitudinal profile\\ 
%- the reconstructed shower depth at maximum ($X_{max}$) 
%is in the field of view of the telescope\\ 
- minimum observed depth $<$ $X_{max}$ $<$ maximum observed depth\\
- a reconstructed longitudinal profile wider than 200 g cm$^{-2}$.
\begin{floatingfigure}[r!]{6.9cm}
    \includegraphics[width=6.5cm]{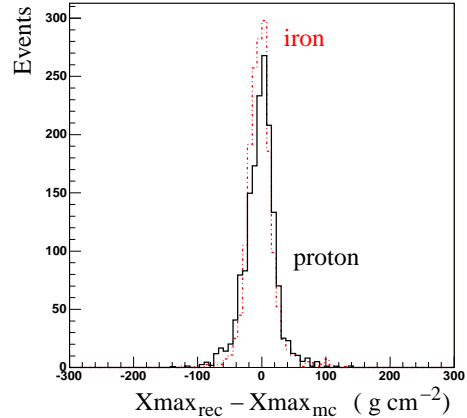}
\protect\caption{\label{fig:xmaxresolution} Residual distribution of the reconstructed   
depth at shower maximum (reconstructed $X_{max}$ - true $X_{max}$) - fixed geometry -,  for 
proton (black line) and iron (red dot-dashed line).}
\end{floatingfigure}
Fig.~\ref{fig:energy-resolution}  
shows the residual distribution  of the reconstructed energy, 
(reconstructed energy-true energy)/(true energy), for events with fixed geometry (blue line)
and with reconstructed geometry (red dot-dashed line). 
The energy resolution improves  
 from 11\% to  9\% in terms of RMS  and the number of selected events
 increases - by a factor 2 -  
 if the geometry 
is set to the true value.
This demonstrates how much the performance of the
 Auger Observatory can improve when operating in hybrid mode.   
The RMS of the residual for the case of fixed geometry 
is shown in Fig.~\ref{fig:energy-RMS} 
as a function of true energy (left), zenith angle (middle) 
and core distance (right).
It depends weakly on the energy (improving slightly with increasing energy)   
and has a stable average value 
of about 9\% over the studied core distance and zenith angle range.
The energy resolution shown has been calculated for 
proton primaries and for a clean atmosphere
(aerosol horizontal attenuation length at sea level of 24 km and scale height of 2 km). 
A test performed with a different atmosphere 
(aerosol horizontal attenuation length at sea level of 12.5 km and scale height of 2 km) shows 
that the energy resolution may degrade from 11\% to 13\% 
for the proton case. \\
Finally, the residual of the reconstructed   
depth at shower maximum (reconstructed $X_{max}$ - true $X_{max}$) 
is shown for fixed geometry in Fig.~\ref{fig:xmaxresolution}
for proton (black line, RMS=25 g cm$^{-2}$) and iron (red dot-dashed line, RMS=22 g cm$^{-2}$).

\section{Comparison with experimental data}

\begin{figure}
\begin{center}
\epsfig{file=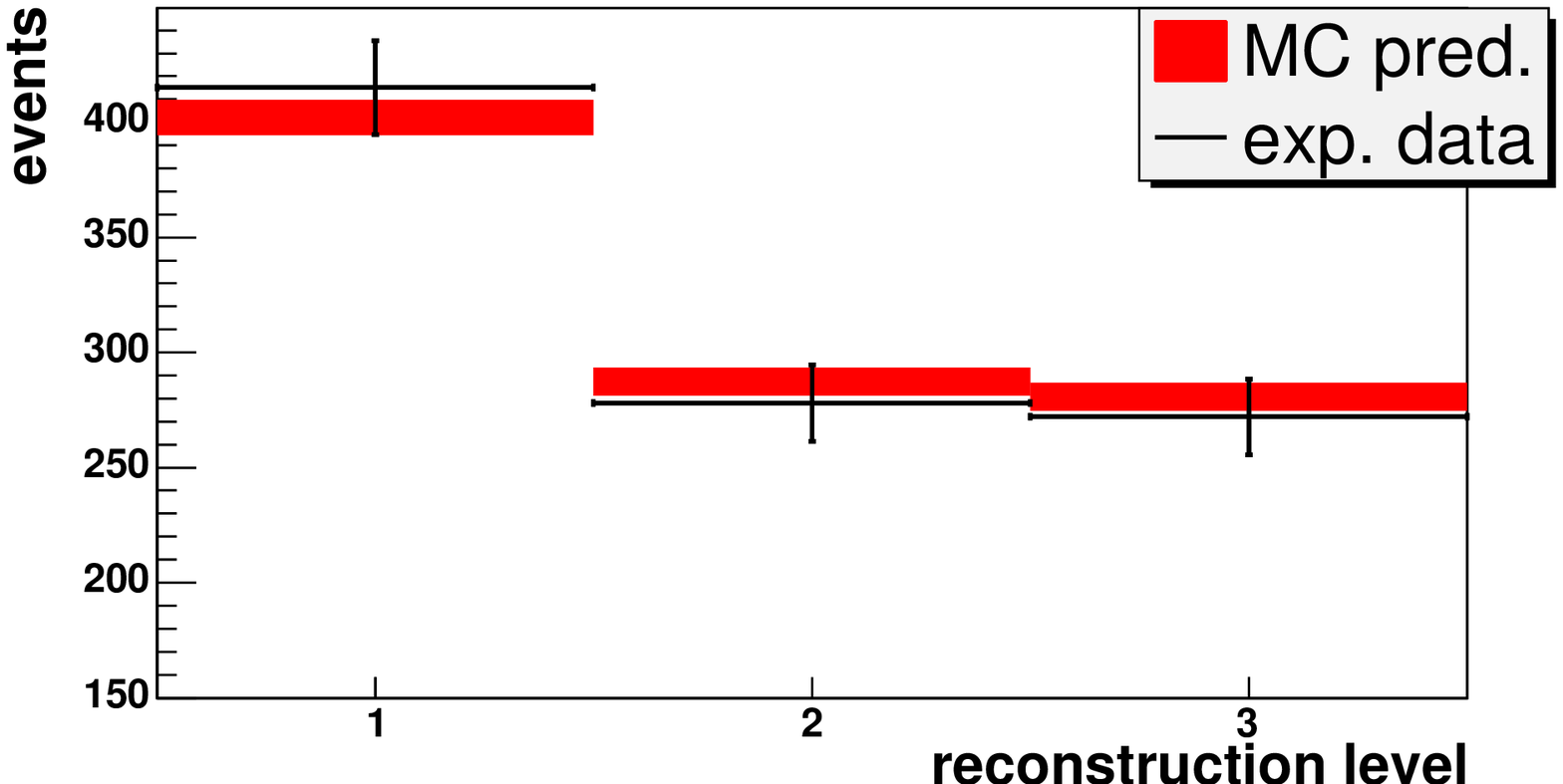,width=7.5cm}
\epsfig{file=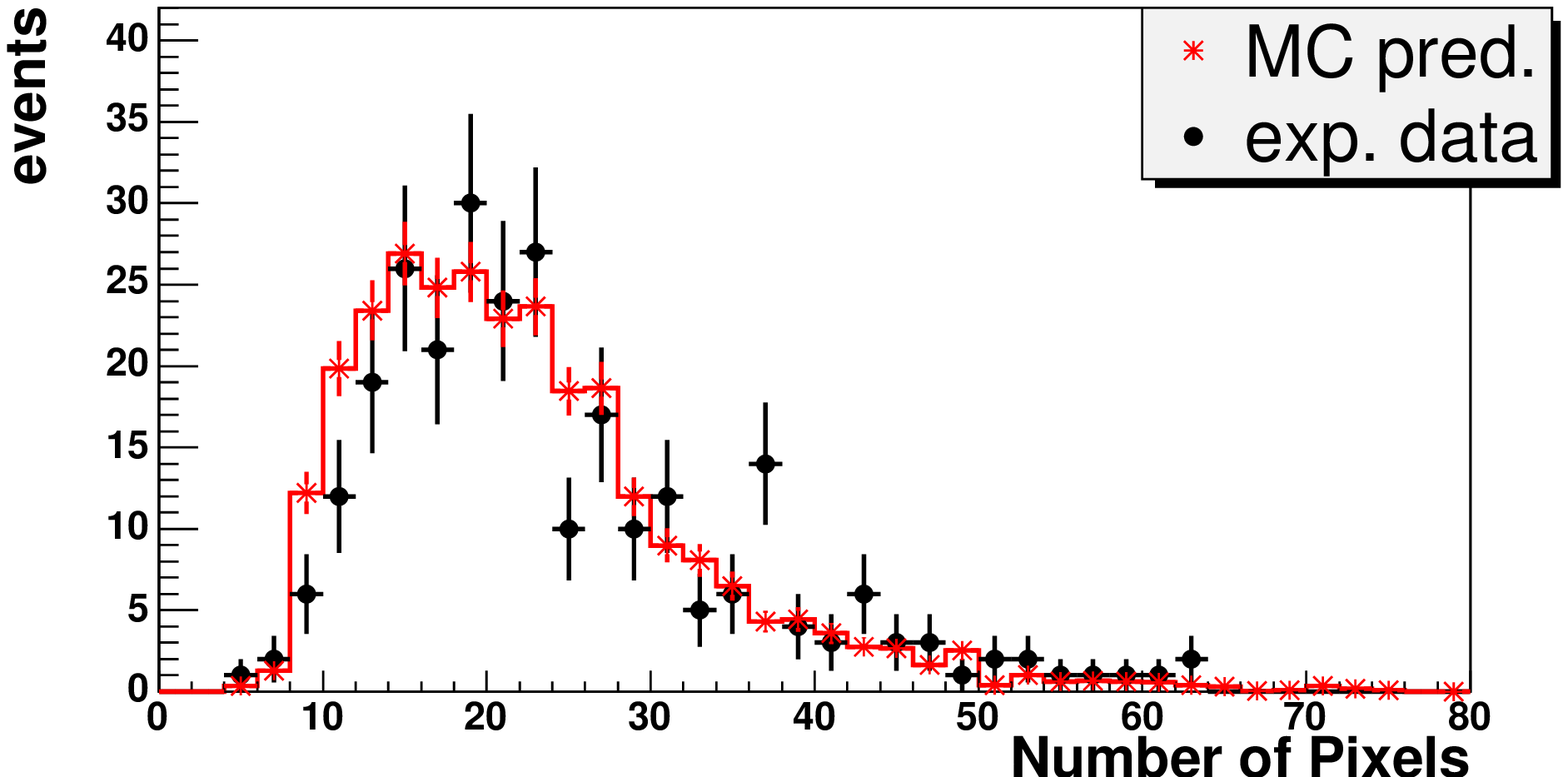,width=7.5cm}
\caption{ Left: Expected number of events (after weighting) (red boxes) 
and experimental data (black line) at different 
  reconstruction levels:  
  geometry reconstructed 
 (level 1) profile reconstructed (level 2) and physical cuts as described 
 in section~\ref{energyres} applied (level 3).  
  The Monte Carlo prediction includes 
  the estimated uncertainty on livetime. Only statistical errors 
  are given for data. 
  Right: Predicted and experimental data distribution of pixels used for
   the longitudinal profile reconstruction.}
\label{Mcdata}
\end{center}
\end{figure}

The trigger simulation has been validated by comparing the predicted number of triggered 
events to the experimental data. 
The Monte Carlo sample has the following characteristics:\\
- 250000 events (50\% iron, 50 \% proton primaries),
 analytical shower profiles (Gaisser-Hillas functions)\\
- energy spectrum generated from $10^{17.5}$eV up to $10^{20.5}$eV according
  to a power-law spectrum with differential spectral index -2\\
- zenith angles generated according to $dN/d\!\cos\theta \propto
  \cos\theta$ between 0$^\circ$ and 60$^\circ$\\
- events simulated in the field of view of Bay 4 of the Los
  Leones eye with landing points distributed uniformly on the surface.\\
To make the Monte Carlo sample comparable with the experimental data, the events have
been re-weighted according to particular physical assumptions.
 This study used a power law cosmic ray spectrum with a break at $10^{18}$ eV
 ($\gamma$= -3.3 for E $< 10^{18}$ eV and $\gamma$= -3 for E $> 10^{18}$ eV,
  motivated by ~\cite{hires})
  and with isotropically distributed arrival directions.
  Two months of data (Los Leones, Bay 4, August-September 2004) have been used, 
  with an estimated total livetime of  $T_l=708000s \pm 10$\%.
The expected number of events has been compared 
with the collected data 
at different levels: geometry reconstructed 
 (level 1), profile reconstructed (level 2) and after applying physical cuts as described 
 in section~\ref{energyres} (level 3). 
%\begin{itemize}
%\item Level 1: requires a reasonable geometry reconstruction that is defined
%  by $\frac{\chi^2_{TimeFit}}{NDOF}<5$
%\item Level 2: requires a reasonable shower profile reconstruction that is
%  defined  by $\frac{\chi^2_{ProfileFit}}{NDOF}<5$
%\item Level 3: Consists of above described physical cuts  
%\end{itemize}
The result is plotted in Fig.~\ref{Mcdata} (left) as a function 
of the reconstruction level; the red boxes show the Monte Carlo 
prediction and the black histogram the experimental data.
The uncertainty on the Monte Carlo expectation includes statistical and systematic 
errors, dominated by the estimated uncertainty on the total livetime; 
only statistical errors are given for data.
The experimental data and prediction agree at each reconstruction level  
within their uncertainties, demonstrating that the trigger simulation is
well understood.
As an example of the detailed consistency between data and simulation, the distribution of the 
number of pixels used for the longitudinal profile reconstruction is shown 
in Fig.~\ref{Mcdata} (right) at the last reconstruction level.
%The number of predicted events (after weighting) and  
%and real data  are shown 
% in Fig.~\ref{Mcdata} (right side).  
%Moreover one can look at the predicted and experimental measured number of
%triggered pixels from the feature extraction. The differential distribution at
%final level is given in figure \ref{Mcdata}. One can see that the MC and
%experimental data agree within shape and normalisation within their
%uncertainties.

%%%%%%%%%%%%%%%%%%%%%%%%%%%%%%%%%%%%%%%%%%%%%%%%%%%%%%%%%%%%%%%%%%%%%%%%%%%%%%%

%\begin{figure}
%\begin{center}
%\epsfig{file=icrc_figures/Bay4PassRateICRCW4.eps,width=7cm}
%\epsfig{file=icrc_figures/PulsPixel.L2.eps,width=8.3cm}
%\caption{ Left: Predicted number of events at each
%  reconstruction level and experimental data. The red box is an estimate of
%  systematic uncertainties mostly due to the livetime. Right: Predicted distribution of triggered pixels and
%  distribution from experimental data.}
%\end{center}
%\label{Mcdata}
%\end{figure} 
%\begin{table}
%\begin{center}
%\begin{tabular}{c|c|c|c}
%&&MC prediction&exp. data\\\hline
%level 1&geometry rec. fine&402.17&415\\\hline
%level 2&profile rec. fine &287.36&278\\\hline
%level 3&physical cuts applied   &280.82&272\\\hline
%\end{tabular}
%\end{center}
%\label{tab1}
%\caption{Summary table}
%\end{table}

%%%%%%%%%%%%%%%%%%%%%%%%%%%%%%%%%%%%%%%%%%%%%%%%%%%%%%%%%%%%%%%%%%%%%%%%%%%%
\vspace{-3pt}
\section{Conclusions}
\vspace{-3pt}

   The performance of the Auger fluorescence detector has been studied 
 using a large number of simulated CORSIKA showers. 
The energy resolution has been estimated for 
the case of known fixed shower geometry which provides 
a realistic estimate for the hybrid operation of the Pierre Auger Observatory.
In this case, the energy resolution improves   
and the number of reconstructible events is larger
by a factor 2 with respect 
to the case of the pure monocular reconstruction. 
The overall energy resolution depends weakly on the shower energy 
and remains stable with an average  
value (RMS) of 9\% over the studied range of 
zenith angles (0$^\circ$ - 60$^\circ$) and core distances (5 - 60 km).
The overall resolution of the atmospheric depth at shower maximum 
is at the level of 22 g cm$^{-2}$ for iron and 25 g cm$^{-2}$ for protons.\\   
Finally, a comparison between simulation and data 
has been carried out at trigger level
for a large sample of analytical shower profiles (Gaisser-Hillas functions); 
agreement has been observed at each reconstruction level.

%%%%%%%%%%%%%%%%%%%%%%%%%%%%%%%%%%%%%%%%%%%%%%%%%%%%%%%%%%%%%%%%%%%%%%%%%%%%
%\section{References}


\vspace{-3pt}
\begin{thebibliography}{99}

%%%%%%%%%%%%%%%%%%%%%%%%%%%  sec 1
\vspace{-5pt}

\bibitem{corsika}
D. Heck \etal,
%{\it ``CORSIKA: A Monte Carlo Code to Simulate Extensive Air Showers''},
Report FZKA 6019, (1998).

\bibitem{fdsim}
L. Prado Jr. \etal,
%{\it To appear in Nuclear Instruments and Methods A}
Nucl. Instr. Meth. A 545, 632 (2005)

\bibitem{offline}
S. Argir\`o \etal, for the Pierre Auger Collaboration, at this Conference, 
usa-paul-T-abs1-he15-poster.

\bibitem{miguel}
M. Mostaf\`a for the Pierre Auger Collaboration, at this Conference. 
usa-mostafa-M-abs1-he14-oral.


\bibitem{bellido}
J. Bellido for the Pierre Auger Collaboration, at this Conference,
aus-bellido-J-abs1-he14-oral.

\bibitem{petrera}
S. Petrera \etal, for the Pierre Auger Collaboration, at this Conference, 
ita-ptrera-S-abs1-he15-poster.

\bibitem{hires}
A. Zech for the HiRes Collaboration, Nucl. Phys. B (Proc. Suppl.) 136 (2004)

\end{thebibliography}
\end{document}